\documentstyle[12pt]{article}

\newtheorem{theoreme}{{\sc Theorem}}[section]
\newtheorem{proposition}[theoreme]{{\sc Proposition}}

\newtheorem{definition}[theoreme]{{\sc Definition}}

\font\twelve=cmbx10 at 15pt
\font\ten=cmbx10 at 12pt

\begin{document}

\begin{titlepage}

\begin{center}

{\ten Centre de Physique Th\'eorique - CNRS - Luminy, Case 907}

{\ten F-13288 Marseille Cedex 9 - France }

{\ten Unit\'e Propre de Recherche 7061}

\vspace{1 cm}

{\twelve A SOLVABLE NONLINEAR}

{\twelve REACTION-DIFFUSION MODEL}

\vspace{0.3 cm}

{\bf Max-Olivier HONGLER}\footnote{Bibos and
Fakult\"at f\"ur Physik, Universit\"at Bielefeld, Permanent address:
Institut de Microtechnique, E.P.F.L., CH 1015 Lausanne, Switzerland}
{\bf and Ricardo LIMA}

\vspace{3 cm}

{\bf Abstract}

\end{center}

We construct a coupled set of nonlinear reaction-diffusion equations
which are exactly solvable. The model generalizes both the Burger
equation and a Boltzman reaction equation recently introduced  by
Th.~W.~Ruijgrok and T.~T.~Wu.

\vspace{3 cm}

\noindent Key-Words : non-linear dynamics, reaction-diffusion,
solvable model.

\bigskip

\noindent October 1993

\noindent CPT-93/P.2956

\bigskip

\noindent anonymous ftp or gopher: cpt.univ-mrs.fr

\end{titlepage}

\section{Introduction}

Despite the enormous effort done since longtime, there are only few
exactly solvable nonlinear field equations in $(1+1)$ dimensions.
Among these models, the Burger equation is a well known example of
diffusion nonlinear equation.

In their remarkable contribution [1], Th. W. Ruijgrok and T. T.
Wu presented a completely solvable model of a
discrete velocity, non-linear Boltzman equation (here referred as
the RW equation). Basically, the non-linear
evolution can be linearized via a logaritmic transformation of
the Hopf-Cole type. In that sense, the RW model has been interpreted
to be a generalization of the Burger equation [2]. Indeed, the Burger
equation describes the evolution of the velocity field of a transport
process governed by the White noise process (WNP), [3], while the RW
equation can be viewed similarly as the evolution for the coupled
velocity fields of a transport process governed by  the random
telegraph process, (RTP) [2].

In this paper we present a completely solvable nonlinear
reaction-diffusion model. The starting point are the
Chapman-Kolmogorof equations corresponding to a dynamical system
driven by a sum of the WNP and the RTP. The nonlinear equations
obtained after an inverse Hopf-Cole transformation are a couple set of
nonlinear reaction-diffusion equations. The exact solution of this
dynamics  is expressible in terms of a linear differential operator
acting on the logarithm of the convolution product of a Gaussian with
the solution of the telegrapher's equation. In this way it is possible
to construct all the solutions corresponding to initial conditions in
the physical domain, i.e. such that they  remain positive, as it might
be for a distribution field. The model [1] and its solution is
 restituted when one of
the control parameters (i.e. the diffusion constant governing the
WNP) is set equal to zero.

\section{The model and its solution} Consider the coupled set of
non-linear reaction-diffusion equations :
\begin{equation} {\partial \over \partial t} + v{\partial \over
\partial x} f_{1}(x,t) =
\mu {\partial^{2} \over \partial x^{2}} f_{1}(x, t) + K_{\mu} (f_{1},
f_{2}, {\partial \over
\partial x}f_{1}, {\partial \over \partial x}f_{2})
\end{equation}
\begin{equation} {\partial \over
\partial t} - v{\partial \over \partial x} f_{2}(x, t) = \mu
{\partial^{2} \over \partial x^{2}}f_{2} (x, t) - K_{\mu}( f_{1},
f_{2}, {\partial \over \partial x }f_{1}, {\partial \over \partial
x}f_{2})
\end {equation}
with the non-linear operator $K_{\mu}$ defined as :
\begin{equation}
\begin{array}{ll}
K_{\mu} = &-\alpha f_{1} +
\beta f_{2} + f_{1} f_{2}\\
&\\
&+{\mu S\over 4v}\left[{(\alpha+\beta)\over v}S-4D_x+{SD\over v}
\right]\\
&\\
&+\mu^2\left[-{S^4\over 16v^4}+{1\over 2}{S^2Sx\over v^3}-
{(S^2)x\over 2v^2}\right]\\
\end{array}
\end{equation}

where $\alpha,\beta,v,\mu$ are positif reals, $S=f_1+f_2$ and
$D=f_1-f_2$. Subscript $x$ denotes the derivative with respect to $x$
and the arguments $(x,t)$ of $f_1$ and $f_2$ have been omitted in the
definition of $K_{\mu}$ for simplicity.

\noindent
\underline{Remark}. When $\mu = 0$, the model defined by Eqs. (1),
(2) and (3) reduces to the RW equation [1]. In this case the model
has the following properties: (1)
conservation of the number of particles and energy; (2) nonlinearity;
(3) positivity of distribution functions and (4) uniqueness of the
equilibrium distribution for any given density.

In our case diffusion clearly prevents property (1) and property (4)
may only be verified with respect to each initial condition. The
remaining properties are still verified in our case.

\begin{definition}
We define the following domain of admissible initial conditions:
\begin{equation}
\begin{array}{ll}
\Gamma(\alpha,\beta,\mu,v)=&\left\{f_1\times f_2\in C^2({\bf R})
\times C^2({\bf R})\vert D(x,0)\right.\\
&\\
&\left.-{\mu\over\nu} S_x(x,0)-{\mu\over 2\nu^2}S^2(x,0)+\alpha
+\beta\right\}
\end{array}
\end{equation}
\end{definition}

We denote by $G(x,t)$ the usual density for the WNP, namelly
\begin{equation}
G(x,t)={1\over \sqrt{2\pi\mu t}}\exp \left\{{x^2\over \mu t}\right\}
\end{equation}
and by $T(x,t)$ the corresponding density for the RTP, namelly
\begin{equation}
T(x, t) = W(x, t) {\rm exp} \lbrace - {\alpha + \beta \over 2}t
- { \alpha
- \beta \over 2v }x \rbrace
\end{equation}
where
\newpage
$$
W(x, t) = {1 \over 2} \left [W_{0} (x + vt) + W_{0} (x-vt) \right ] +
$$
$$+ {1 \over 2} \int_{x -vt} ^{x+vt} I_{0} \left ( { \sqrt{\alpha
\beta}
\over v} ( v^{2} t^{2} - (x - x')^{2} )^{{1 \over 2}} \right ) W_{1}
(x') dx' $$
\begin{equation}
+ {1 \over 2}\sqrt{\alpha\beta}t \int _{x-vt}^{x+vt}\left ( { I_{1}
\left (  {\sqrt{\alpha \beta} \over v}
(v^{2} t^{2} - ( x - x')^{2} )^{{1 \over 2}}\right ) \over
\sqrt{v^{2} t^{2} - (x-x')^{2}} }
W_{0}(x')
\right) dx'
\end{equation}
where $I_0$ and $I_1$ denote modified Bessel functions and
\begin{equation}
W_{0}(x) = {\rm exp} \left\{ -
{1 \over 2}\left[ \int^{x} S(x', 0)dx' +  { \alpha -
\beta \over v}\right] \right\}
\end{equation}
and
\begin{equation}
W_{1}(x) = {1 \over 2} \left\{ D(x, 0) - {\mu\over\nu}S_x(x, 0) -
{\mu
\over2\nu^2}S^2(x, 0) +
\alpha + \beta
\right\} W_{0}(x)
\end{equation}
The next proposition is the main result of this paper.

\begin{proposition}
The solution of the set of reaction-diffusion equations (1), (2) for
initial conditions
$\left ( f_{1}(x, 0), f_{2}(x, 0) \right ) \in \Gamma$, reads
\begin{equation}
f_{1}(x, t) = {\cal O}^{-} {\rm ln} \left( N(x, t) \right)
\end{equation}
\begin{equation}
f_{2}(x, t) = {\cal O}^{+} {\rm ln} \left( N(x, t) \right)
\end{equation} with the operators ${\cal O}^{\pm}$ defined by :
\begin{equation}
{\cal O}^{\pm} = \left ( {\partial \over \partial t}
\pm v{\partial
\over \partial x} - \mu {\partial^{2} \over \partial x^{2}} \right )
\end{equation} and
\begin{equation}\label{conv}
N(x, t) =\int_{R}^{} G(x', t)T(x-x',t)dx'
\end{equation}
\end{proposition}

{\sc Proof}
We follow the idea of the construction presented in [1]. Hence, let
the operators
${\cal O}^{\pm}$ as defined in Eq.(12). These operators commute.
Therefore it exists a field $F(x, t)$ such that Eqs. (1) and (2) are
equivalent to :
\begin{equation}
{\cal O}^{+} {\cal O}^{-} F(x, t) = - K_{\mu} \left ( - {\cal O}^{-}
F,  {\cal O}^{+} F, -{\cal O}^{-} {\partial \over \partial x} F, {\cal
O}^{+}  {\partial \over \partial x}
F\right ) \end{equation} with
\begin{equation}
-{\cal O}^{-} F(x, t) = f_{1}(x, t)
\end{equation}
and
\begin{equation}
{\cal O}^{+} F(x, t) = f_{2}(x, t)
\end{equation}
Consider now the field $ N(x, t) : (x, t) \in
R \times R^{+} \rightarrow R^{+}$
defined by :
\begin{equation}
N(x, t) = {\rm exp} \lbrace - F(x, t) \rbrace
\end{equation}
Introducing Eq. (17) into Eq. (14), after some tedious manipulations,
it is possible to verify that
$f_1$, $f_2$ will satisfy the system  (1) - (2) with $K_{\mu}$
defined as in (3) provided the field
$N(x,t)$ obeys the following evolution equation :
\begin{equation}
{\rm det} \left ( \matrix{-\lbrack \alpha + {\cal O}^{+} \rbrack &
\beta\cr \alpha & -\lbrack \beta + {\cal O}^{-} \rbrack \cr} \right )
N(x, t) = 0
\end{equation}
Eq. (18) is a fourth
order hyperbolic partial differential differential equation with a
 probabilistic interpretation, [4, 5]. Consequently, Eq. (18)
conserves the positivity of its solution. This equation has been first
encounterd in [4]  (see Eq. (21) of this reference) and its solution
is given in [5], where it is shown that :
\begin{equation}
N(x, t) = \left ( G * T \right) (x, t)
\end{equation}
with $*$ denoting the convolution product as in Eq.(13) with :
\begin{equation}
{\partial \over \partial t} G(x, t) = \mu^{2} {\partial^{2} \over
\partial x^{2}} G(x, t)
\end{equation}
and
\begin{equation}
\left ( {\partial^{2} \over \partial t^{2}} + (\alpha + \beta)
{\partial \over
 \partial t}\right ) T(x, t) =\left ( v^{2} {\partial^{2} \over
\partial x^{2}} +  v (\alpha - \beta)
{\partial \over \partial x} \right ) T(x, t).
\end{equation}
{}From Eqs. (10), (11) and (12), we have :
\begin{equation}
N(x, 0) = N_{0}(x) = {\rm exp} \lbrace -{ 1 \over 2v} \int dx \left
( f_{1} (x, 0) +
f_{2}(x, 0) \right )\rbrace
\end{equation}
and
\begin{equation}
{\partial \over \partial t} N(x, 0) = N_{1}(x) = {1 \over 2} \lbrace
f_{1}(x, 0)
 - f_{2}(x, 0) +{2 \mu\over v} {\partial \over \partial x}\lbrack
f_{1}(x, 0) +
f_{2}(x, 0)\rbrack \rbrace
N_{0}(x).
\end{equation}
In order to get $G(x,t)$ as in Eq.(5) it is sufficient to start with
the initial value in the form :
\begin{equation}
G(x,0)=\delta(x)
\end{equation}
Now, concerning $T(x,t)$, in order to match Eq. (6.4) of [1] we
introduce $W(x,t)$ defined by Eq. (6). In view of (19) and (6)
$N(x,t)$ will be known as far as the solution $W(x,t)$ of the
following equation is found :
\begin{equation} {\partial^{2} \over
\partial t^{2}} W(x, t) = \left ( v^{2} {\partial^{2}  \over \partial
x^{2}} + \alpha \beta \right) W(x, t) \end{equation} To integrate (25)
one needs initial conditions. The last are related to the initial
conditions for
$N(x,t)$ by
\begin{equation}
W(x, 0) = W_{0} (x) = {\rm exp}
\lbrace \left ( {\alpha - \beta \over 2 v} \right) x \rbrace N_{0}(x)
\end{equation}
and hence with the initial conditions for $f_1$ and $f_2$. Therefore
they may be computed using (19)-(24) which leads to (8) and (9).

Finally, thanks to Eqs.(8), (9) and (26), the solution of Eq. (25) is
given in [1] and reads as written in Eq. (7). For the initial values
$\left (f_{1}(x, 0), f_{2}(x, 0) \right) \in \Gamma$ as defined by Eq.
(4), $W_{0}(x)$ and $W_{1}(x)$ are positive. Hence $W(x, t ) \geq 0
,\forall (x, t)
\in R \times R^{+}$, see Lemma 1.3.2 in [6]. Hence $N(x, t) $ is
positive since by Eq.(19) it is the convolution of two positive
fields. Therefore ${ \rm ln}
\left ( N(x, t) \right) $ is well
defined. The proof is complete.

\section{Concluding remarks}
In this note, we have shown how to extent the exactly solvable
discrete  velocity non-linear Boltzman equation introduced by Th. W.
Rujgrok and T. T . Wu [1], (the RW equation). In this way we obtain a
nonlinear reaction-diffusion coupled system of equations for which the
exact solutions can be computed as far as the corresponding initial
conditions are in the positivity preserving domain. Our construction
is based on the observation that that the RW equation and the Burger
equation are non-linear evolutions for velocity fields of linear
transport processes, a point of view pioneered in [4] and further
extended in [2]. Accordingly, the construction presented here can be
further generalized by considering sets of non-linear equations for
velocity fields describing a transport process with dynamics governed
by a mixed noise  stochastic differential equations [5]. Therefore,
the model presented here is only one of the simplest of a whole family
to be further investigated.

{}From that perspective, the special case sorted out here might appear
to be an arbitrary choice. However, due to to the simple form of its
solutions and its reaction-diffusion form, it may deserve special
attention.

In a forthcoming paper we shall study some properties of the
solutions of this model.

\end{document}